\documentclass{bio}
\usepackage[T1]{fontenc}
\usepackage[utf8]{inputenc}
\usepackage{url}
\usepackage{float}
\makeatletter
\def\ps@plain{%
  \def\@oddhead{}%
  \def\@evenhead{}%
  \def\@oddfoot{}%
  \def\@evenfoot{}%
}
\def\ps@headings{%
  \def\@oddhead{}%
  \def\@evenhead{}%
  \def\@oddfoot{}%
  \def\@evenfoot{}%
}
\makeatother

\begin{document}
\vspace*{-4cm}

\title{Spatio-Temporal Log-Gaussian Cox-Hawkes Processes with Inhibition and Excitation for Stochastic Star Formation}

\author{Qihan Zou \\[4pt]
\textit{University of Melbourne}
\\[2pt]
{qihanzou@alumni.unimelb.edu.au}}

\maketitle
\thispagestyle{empty}
\begin{abstract}
{We establish a connection between the stochastic self-propagating star-formation (SSPSF) model and spatio-temporal point processes by showing that the SSPSF update law admits a conditional Poisson representation. Building on this connection, we propose a spatio-temporal log-Gaussian Cox-Hawkes process as a continuous point process model for stochastic star formation. The model represents star-formation events as point patterns driven jointly by deterministic galactic structure, latent spatio-temporal background variation, and dependence on past events. Its key feature is that the deterministic mean field, latent Gaussian random field, and history-dependent interaction field enter through a single log-intensity. This log-scale construction differs from additive Cox-Hawkes formulations and allows the history effect to be signed, past events may either increase or decrease future local intensity while the conditional intensity remains positive. The resulting framework provides an interpretable point-process model for representing latent clustering, self-excitation, local inhibition, and event-driven propagation in stochastic star formation. Further extensions replace the constant history coefficient with deterministic or random fields representing source influence, local response or both. Beyond linking SSPSF to spatio-temporal point-process theory, it offers a continuous stochastic formulation for analysing the propagation of star formation in galaxies and for interpreting observational surveys of star-forming regions within a unified statistical model.
}
\end{abstract}
galaxy formation; galaxy evolution; Cox process; Hawkes process; self-excitation

\section{Introduction}
The stochastic self-propagating star formation (SSPSF) model, originating in the self-propagating star-formation model of \citet{mueller1976propagating} and developed in stochastic form by \citet{gerola1978stochastic, gerola1980theory}, provides a probabilistic framework for modelling the formation and propagation of star-forming regions across galactic disks. Unlike traditional density-wave explanations of spiral structure, such as those discussed by \citet{lin1979density, lin1987spiral, shu2016six}, the SSPSF framework explains a range of galactic morphologies through stochastic local triggering, feedback, and differential rotation. In this model, newly formed massive stars or star-forming regions may induce subsequent star formation in neighbouring regions, leading to clustered and self-propagating stellar patterns. Several extensions have refined the original SSPSF mechanism. \citet{comins1981profiles} allowed the triggered star-formation probability to vary with galactic radius, representing the decline of star-formation efficiency in the outer disk. \citet{statler1983stochastic} extended the model to three dimensions by stacking multiple two-dimensional layers and adding vertical neighbour connections. Percolation-based extensions proposed by \citet{seiden1990percolation} further emphasised the importance of local connectivity and threshold-induced activation. Later, \citet{jungwiert1994stochastic} introduced anisotropic propagation rules, in which the triggering probability is distributed elliptically around a star-forming cell, with the orientation and eccentricity linked to the galactic rotation curve. Together, these developments show that SSPSF models provide a flexible rule-based framework for studying how local interactions, stochastic triggering, and feedback processes can generate large-scale star-formation patterns. The SSPSF framework is also conceptually related to sandpile-type models of self-organised criticality. The Abelian sandpile model, introduced by \citet{bak1987self} and later developed analytically by \citet{dhar1990self}, shows how simple local rules, threshold activation, and redistribution can generate complex cascading behaviour without global control. In this sense, sandpile models provide a useful analogue for SSPSF, since both frameworks describe how local activation may propagate through neighbouring regions and produce large-scale spatial structure.

Spatio-temporal point processes provide a natural framework for analysing events observed in both space and time \citep[see e.g.,][]{ogata1988statistical, daley2003introduction, daley2008introduction, diggle2006spatio, diggle2013statistical}. Building on general point process theory \citep{cox1980point}, early work developed explicit space-time models for applications such as forest dynamics \citep{rathbun1994space} and introduced second-order methods for detecting space-time interaction and clustering \citep{diggle1995second}. These developments established spatio-temporal point processes as flexible tools for studying event patterns shaped by spatial structure, temporal variation, and event history \citep{diggle2006spatio,gonzalez2016spatio}. Self-exciting models are widely used for clustered event data. The Hawkes process was introduced as a temporal self-exciting point process by \citet{hawkes1971spectra}, with its cluster representation developed by \citet{hawkes1974cluster}. It was later extended to spatio-temporal settings, particularly through earthquake models and related branching formulations \citep[e.g.,][]{ogata1988statistical,ogata1998space,zhuang2002stochastic,veen2008estimation}. These models decompose event risk into a background component and a history-dependent triggering component, thereby separating exogenous variation from events induced by past events. This interpretable structure has led to broad use in earthquake modelling, crime forecasting, disease modelling, and other clustered event applications \citep[e.g.,][]{reinhart2018review, bernabeu2025spatio}. However, classical spatio-temporal Hawkes models are mainly excitatory, since past events are assumed to increase future event risk. They are therefore less suitable when previous events may suppress occurrence nearby in space and time. Several recent works have extended Hawkes processes beyond the purely excitatory setting. \cite{malem2021nonlinear} considered non-linear Hawkes processes with Gaussian process self-effects, allowing past events to have either excitatory or inhibitory effects. \cite{bielecki2025loglinear} studied log-linear Hawkes processes, where the rate function is exponential and the conditional intensity is a log-linear function of past events. \cite{bonnet2023inference} developed inference methods for multivariate exponential Hawkes processes with inhibition. These works study signed or non-linear Hawkes-type dynamics, but they do not consider a model in which a latent spatio-temporal log-Gaussian Cox field and a signed Hawkes-type history field are combined within a single log-intensity. On the other hand, Log-Gaussian Cox processes \citep[see e.g.,][]{baddeley2016spatial, moller2017some} provide a complementary framework for modelling clustering through latent background variation. Introduced by \citet{moller1998log} as Cox processes with log-Gaussian random intensities, they were later extended to spatio-temporal prediction with stochastic variation over both space and time \citep{brix2001spatiotemporal}. Conditional on the latent intensity surface, events follow an inhomogeneous Poisson process, with the log intensity usually represented by covariates and a latent Gaussian field \citep{diggle2013statistical,diggle2013spatial}. These models are well suited to clustering driven by unobserved background field \citep{moraga2023spatial}, but they do not explicitly model event interaction and therefore cannot directly separate latent clustering from triggering by previous events. A closely related development is the Cox-Hawkes model, which combines a stochastic Cox process background with Hawkes-type triggering in order to separate latent spatio-temporal variation from event-induced excitation \citep{miscouridou2022cox}. However, because the triggering component is usually additive and nonnegative, such models primarily capture excitation and do not naturally represent inhibitory history effects. 

While the classical stochastic self-propagating star formation model and spatio-temporal point processes have both been widely studied, their relationship has not been explicitly formulated. For each eligible cell, we combine spontaneous star formation and triggering by neighbouring active cells in a conditional Poisson construction, with a positive count producing a single activation. The detailed description of the classical SSPSF model and the conditional Poisson representation of the SSPSF update law are provided in Appendix~\ref{app:sspsf-stpp}. Building on this connection and motivated by the limitations of additive Cox-Hawkes formulations \citep{miscouridou2022cox}, we then propose a new continuous stochastic model for star formation: a spatio-temporal log-Gaussian Cox-Hawkes process with signed history dependence. The proposed model combines deterministic galactic structure, latent spatio-temporal background variation, and history-dependent interaction within a single log-intensity. Unlike additive Cox-Hawkes models, the history effect enters on the log-intensity scale and can therefore be either excitatory or inhibitory while the conditional intensity remains positive. Thus, the model is not simply a continuous version of the classical SSPSF model, but a new point-process framework for modelling stochastic star formation with latent background variation, self-excitation, local inhibition, and event-driven propagation. While our existence and explosion arguments are inspired by the Poisson embedding approach for log-linear Hawkes processes studied by \citet{bielecki2025loglinear}, the present model has a different purpose and structure. To account for differences in the effects of past events, we allow the history coefficient to depend on the location and time of a past event, the location and time at which its influence is evaluated, or both. These extensions also allow the coefficient fields to be random.

The remainder of this work is structured as follows. Section 2 introduces the proposed spatio-temporal log-Gaussian Cox-Hawkes process with signed history dependence. Section 3 discusses its basic properties and special cases. Sections 4 and 5 study existence, non-explosion, and possible explosion of the proposed process. Section 6 introduces the recovery-time version of the model and proves its non-explosion. Section 7 gives a physical interpretation of the recovery-time model for stochastic star formation. Section 8 presents extensions with deterministic or random interaction coefficient fields. Section 9 concludes with a discussion and outlook. Appendix~\ref{app:sspsf-stpp} provides the detailed background on the classical SSPSF model and the conditional Poisson representation of the SSPSF update law.

\section{Log-Gaussian Cox-Hawkes Process}
Let $\mathcal{S}\subset\mathbb{R}^2$ be a bounded spatial domain and let $[0,T]$ be the observation window. We observe a spatio-temporal point pattern $\mathcal{W}=\{(s_i,t_i):i=1,\ldots,n\}$, where $s_i\in\mathcal{S}$ and $0<t_1<...<t_n\leq T$. Let $\mathcal{H}_t=\{(s_i,t_i):t_i<t\}$ denote the history before time $t$. Conditional on the latent field $Y$, we model the process through its conditional intensity $\lambda(s,t\mid\mathcal{H}_t)$, with conditioning on $Y$ suppressed in the notation, and propose a log-Gaussian Cox-Hawkes process of the form $$\lambda(s,t\mid\mathcal{H}_t)=\exp\{\mu(s,t)+Y(s,t)+D(s,t\mid\mathcal{H}_t)\},$$ where $\mu(s,t)$ is a deterministic mean field, $Y(s,t)$ is a latent spatio-temporal Gaussian random field, and $D(s,t\mid\mathcal{H}_t)$ is a history induced dynamic field. The latent field captures residual spatio-temporal variation, while the dynamic field captures the effect of previous events on future event intensity. We assume that $Y(s,t)$ is a mean zero Gaussian process, $Y(s,t)\sim \mathrm{GP}\{0,C_Y\}$, with separable exponential covariance function $$ C_Y((s_1,t_1),(s_2,t_2)) = \sigma_Y^2 \exp\left\{ -\frac{\|s_2-s_1\|}{\phi_Y} -\theta_Y |t_2-t_1| \right\}. $$ Here $\sigma_Y^2>0$ is the marginal variance, $\phi_Y>0$ controls the spatial correlation range, and $\theta_Y>0$ controls the temporal rate of decorrelation. We assume that $\mu(s,t)$ is Borel measurable and bounded on $S\times[0,T]$ for every finite $T>0$, and use a version of $Y$ with continuous sample paths on $\overline{S}\times[0,\infty)$. The specified covariance admits such a version, so $b(s,t)=\mu(s,t)+Y(s,t)$ satisfies the boundedness condition used below almost surely. The dynamic field is defined as $D(s,t\mid\mathcal{H}_t)=\sum_{i:t_i<t}g(t-t_i,s-s_i)$, where $g(u,h)$ is a causal spatio-temporal interaction function satisfying $g(u,h)=0$ for $u\leq 0$. We use the separable specification $$ g(u,h) = \alpha \exp(-\beta u)\varphi_{\Sigma}(h)\mathbf{1}\{u>0\}, $$ where $\alpha\in\mathbb{R}$ controls the direction and magnitude of the history effect, $\beta>0$ controls temporal decay, and $\varphi_{\Sigma}(h)$ is a Gaussian spatial kernel, $$ \varphi_{\Sigma}(h) = \frac{1}{2\pi|\Sigma|^{1/2}} \exp\left\{ -\frac{1}{2}h^\top\Sigma^{-1}h \right\}, $$ with positive definite covariance matrix $\Sigma$. In this work, we define the spatial covariance matrix in the interaction kernel as $$ \Sigma = \sigma_\Sigma^2 \begin{pmatrix} 1 & \eta_\Sigma\\ \eta_\Sigma & 1 \end{pmatrix}, $$ where $\sigma_\Sigma>0$ controls the marginal spatial scale of the history-dependent interaction and $\eta_\Sigma\in(-1,1)$ controls the correlation between the two spatial coordinate directions. When $\eta_\Sigma=0$, this reduces to the isotropic specification $\Sigma=\sigma_\Sigma^2 I_2$. Hence, $$ D(s,t\mid\mathcal{H}_t) = \alpha \sum_{i:t_i<t} \exp\{-\beta(t-t_i)\} \varphi_{\Sigma}(s-s_i), $$ and the full conditional intensity becomes $$ \lambda(s,t\mid\mathcal{H}_t) = \exp\left[ \mu(s,t) + Y(s,t) + \alpha \sum_{i:t_i<t} \exp\{-\beta(t-t_i)\} \varphi_{\Sigma}(s-s_i) \right]. $$

For a spatio-temporal point process observed on $\mathcal{S}\times[0,T]$, the conditional log likelihood associated with the intensity $\lambda(s,t\mid\mathcal{H}_t)$ is $$ \ell = \sum_{i=1}^{n} \log \lambda(s_i,t_i\mid\mathcal{H}_{t_i}) - \int_0^T\int_{\mathcal{S}} \lambda(s,t\mid\mathcal{H}_t)\,ds\,dt. $$ Under the proposed log-Gaussian Cox-Hawkes model, conditional on the latent field $Y(s,t)$, this becomes $$ \begin{aligned} \ell(\Theta\mid Y) &= \sum_{i=1}^{n} \left[ \mu(s_i,t_i) + Y(s_i,t_i) + \alpha \sum_{j:t_j<t_i} \exp\{-\beta(t_i-t_j)\} \varphi_{\Sigma}(s_i-s_j) \right] \\ &\quad - \int_0^T\int_{\mathcal{S}} \exp\left[ \mu(s,t) + Y(s,t) + \alpha \sum_{j:t_j<t} \exp\{-\beta(t-t_j)\} \varphi_{\Sigma}(s-s_j) \right] \,ds\,dt . \end{aligned} $$ Let $\gamma_\mu$ denote the finite dimensional parameters in the deterministic mean field $\mu(s,t)$. We collect the unknown parameters as $ \Theta = \left( \gamma_\mu, \alpha, \beta, \sigma_\Sigma, \eta_\Sigma, \sigma_Y, \phi_Y, \theta_Y \right), $ where $\alpha\in\mathbb{R}$ controls the direction and strength of the history effect, and $\beta>0$ controls its temporal decay. The parameters $\sigma_\Sigma$ and $\eta_\Sigma$ determine the spatial covariance matrix $\Sigma$ in the interaction kernel. The parameters $\sigma_Y>0$, $\phi_Y>0$, and $\theta_Y>0$ determine the covariance structure of the latent Gaussian field, controlling its marginal variability, spatial correlation range, and temporal decorrelation rate, respectively. For fixed $Y$, the conditional log likelihood above does not depend explicitly on $\sigma_Y$, $\phi_Y$, or $\theta_Y$. Inference for these covariance parameters requires accounting for the Gaussian field distribution, for example by integrating the conditional likelihood over $Y$.

\section{Basic Properties and Special Cases}
\subsection{Positivity of the Intensity}
The conditional intensity is defined as$$\lambda(s,t \mid \mathcal{H}_t) =  \exp \{\mu(s,t)+Y(s,t)+D(s,t \mid \mathcal{H}_t)\}.$$ Since the exponential function is strictly positive, the conditional intensity remains positive for all admissible parameter values before the explosion time $T_{\infty}$. This property holds regardless of whether the history-dependent effect is positive or negative.

\subsection{Reduction to a Log-Gaussian Cox Process}
When $\alpha = 0,$ the history-dependent field vanishes, $D(s,t \mid \mathcal{H}_t)=0,$ and the model reduces to
$$\lambda(s,t) = \exp \{\mu(s,t)+Y(s,t)\}.$$ This is the standard spatio-temporal log-Gaussian Cox process. In this case, clustering beyond deterministic intensity variation is entirely driven by the latent Gaussian field.

\subsection{Reduction to a Spatio-Temporal Log-Linear Hawkes Process}
When the latent Gaussian field is absent, $Y(s,t)=0,$ the conditional intensity becomes $$ \lambda(s,t \mid \mathcal{H}_t) = \exp \{\mu(s,t)+D(s,t \mid \mathcal{H}_t) \}. $$ In this case, the model reduces to a spatio-temporal log-linear Hawkes-type process. The dependence on past events enters through the log-intensity rather than additively on the intensity scale. The process no longer contains latent background variation, and local dependence is generated by the history-dependent field.

\subsection{Excitation and Inhibition}
The sign of the parameter $\alpha$ determines the direction of the history effect. When $\alpha > 0,$ previous events increase the future local log-intensity, producing self-excitation. When $\alpha < 0,$ previous events decrease the future local log-intensity, producing local inhibition. When $\alpha = 0,$ the process contains no history dependence and reduces to a log-Gaussian Cox process.

\subsection{Relation to Cox-Hawkes Models}
Cox-Hawkes models typically take the additive form $$ \lambda(s,t\mid\mathcal{H}_t) = \exp\{f(s,t)\} + \sum_{i:t_i<t}g(t-t_i,s-s_i). $$ In these models, the Cox background and the Hawkes triggering term are added on the intensity scale, and the triggering function is usually nonnegative. In contrast, the proposed model places the deterministic mean field, latent Gaussian field, and history-induced dynamic field inside a single log-intensity. As a result, excitation and inhibition can be represented within the same framework, while positivity of the intensity is automatically preserved.

\section{Existence}
Following \cite{bielecki2025loglinear}, this section gives a Poisson embedding construction of the proposed process, conditional on a realisation of the latent Gaussian field. This construction shows that the model is well defined up to its explosion time, $T_{\infty}=\lim_{n\to\infty}T_n.$

The idea is to introduce a background homogeneous Poisson random measure on the three-dimensional space $[0,\infty)\times S\times \mathbb{R}_{\geq 0},$ with coordinates $(t,s,z)$, where $t$ is the time coordinate, $s$ is the spatial coordinate, and $z$ is an auxiliary height coordinate. The Poisson points are spread uniformly in this enlarged space. The proposed point process is then obtained by accepting only those points whose height satisfies $z\leq \lambda(s,t).$ Thus, the Poisson embedding construction turns the intensity function $\lambda(s,t)$ into an acceptance threshold for points from the background Poisson random measure.

\paragraph{Definition 4.1 (log-Gaussian Cox-Hawkes process).}
Let $S\subset\mathbb{R}^2$ be bounded, and let $Y(s,t)=y(s,t)$ be a fixed realisation of the latent Gaussian field. A spatio-temporal point process $N$ on $S\times[0,\infty)$ with explosion time $T_{\infty}\in(0,\infty]$ is called a log-Gaussian Cox-Hawkes process with signed history dependence if its conditional intensity is $$ \lambda^y(s,t\mid\mathcal{H}_t) = \exp\{\mu(s,t)+y(s,t)+D(s,t\mid\mathcal{H}_t)\} \mathbf{1}\{t<T_{\infty}\}, $$ where $$ D(s,t\mid\mathcal{H}_t) = \alpha \sum_{i:t_i<t} \exp\{-\beta(t-t_i)\} \varphi_{\Sigma}(s-s_i). $$ The process is called non-explosive if $T_{\infty}=\infty$ almost surely, and explosive if $T_{\infty}<\infty$ with positive probability.

\paragraph{Lemma 4.2 (Poisson embedding).} 
Let $\mathcal{N}$ be a homogeneous Poisson random measure on $[0,\infty)\times S\times \mathbb{R}_{\geq 0}$ with intensity measure $dt\,ds\,dz$. Thus, the expected number of Poisson points in any measurable region is equal to the Lebesgue measure of that region. Let $(\mathcal{F}_t^{\mathcal{N}})_{t\geq 0}$ be the filtration generated by the Poisson random measure up to time $t$, where $$ \mathcal{F}_t^{\mathcal{N}} = \sigma\left( \mathcal{N}((a,b]\times B\times C): 0<a<b\leq t,\ B\in\mathcal{B}(S),\ C\in\mathcal{B}(\mathbb{R}_{\geq 0}) \right). $$ This $\sigma$-field contains all information from the background Poisson process up to time $t$, namely all candidate points whose time coordinate is at most $t$. Let $ \mathcal{G}_t = \sigma(\mathcal{F}_t^{\mathcal{N}},\mathcal{F}_0^N), \: t\geq0, $ where $\mathcal{F}_0^N$ contains the initial history.

Let $N$ be a spatio-temporal point process on $S\times\mathbb{R}$ with initial condition $N^0$ on $S\times\mathbb{R}_{\leq 0}$, independent of $\mathcal{N}$. Let $$ \mathcal{F}_t^N = \sigma\{N(A):A\in\mathcal{B}(S\times(-\infty,t])\} $$be the natural filtration of $N$. The process $N$ has two parts. The first part is the initial history $N^0$ before time zero. The second part consists of future events on $S\times(0,\infty)$, which will be constructed from the background Poisson random measure $\mathcal{N}$. The independence assumption means that the past history before time zero is independent of the future Poisson randomness used to generate new events. Suppose that the intensity $\lambda(s,t)$ is a nonnegative predictable process with respect to the filtration $(\mathcal{G}_t)_{t\geq0}$. Predictability means that, at time $t$, the value of $\lambda(s,t)$ is determined by information available just before $t$. It cannot depend on future Poisson points. Otherwise, the acceptance rule would be allowed to look into the future, and it would not define a valid point process intensity.

For the future part of the process, assume that $$ N(A) = \int_0^\infty \int_S \int_{\mathbb{R}_{+}} \mathbf{1}\{(s,t)\in A\} \mathbf{1}\{z\leq \lambda(s,t)\} \mathcal{N}(dt,ds,dz), $$ for all $A\in\mathcal{B}(S\times(0,\infty))$. Here, $N(A)$ counts the background Poisson points whose space-time coordinates lie in $A$ and whose auxiliary height $z$ is below the threshold $\lambda(s,t)$. Equivalently, the process $N$ is obtained by accepting candidate points $(t,s,z)$ whenever $z\leq \lambda(s,t).$ Thus, $\lambda(s,t)$ acts as an acceptance surface over space and time. Then $\lambda(s,t)$ is the conditional intensity of $N$ with respect to $(\mathcal{G}_t)_{t\geq0}$. If $N$ is constructed by accepting Poisson points below the predictable threshold $\lambda(s,t)$, then $\lambda(s,t)$ is indeed the conditional intensity of the accepted process. In short, Poisson embedding together with a predictable acceptance rule gives an accepted point process with intensity $\lambda(s,t)$. Furthermore, if $\lambda(s,t)$ is predictable with respect to the natural filtration $(\mathcal{F}_t^N)_{t\geq 0}$ of the accepted process $N$, then $\lambda(s,t)$ is the conditional intensity of $N$ with respect to $(\mathcal{F}_t^N)_{t\geq0}$. There are two information structures. The first one is the larger filtration generated by the background Poisson random measure and the initial condition $N^0$. The second one is the smaller natural filtration generated only by the accepted process $N$. The first part of the lemma shows that $\lambda(s,t)$ is an intensity with respect to the larger filtration. If $\lambda(s,t)$ can be computed from the history of $N$ itself, then it is also an intensity with respect to the natural history of $N$. This is important because the proposed intensity should depend only on the observed history of the process, not on rejected candidate Poisson points.

\medskip

\begin{proof}
Recall that $\mathcal{G}_t=\sigma(\mathcal{F}_t^{\mathcal{N}},\mathcal{F}_0^N)$ is the filtration generated by the background Poisson random measure up to time $t$, and $\mathcal{F}_0^N$ contains the initial history. The first statement follows from the standard compensator characterisation of point processes constructed from Poisson random measures with predictable acceptance functions \citep{bremaud1996stability, bielecki2025loglinear}. The process $N$ is obtained by accepting background Poisson candidate points $(t,s,z)$ whenever $z\leq \lambda(s,t)$. Since the acceptance function is predictable with respect to $(\mathcal{G}_t)_{t\geq0}$, the compensator of the accepted random measure is $\lambda(s,t)\,dt\,ds$. Therefore, for all $B\in\mathcal{B}(S)$ and $0\leq a<b$, $$ \mathbb{E}\{N(B\times(a,b])\mid \mathcal{G}_a\} = \mathbb{E} \left[ \int_a^b \int_B \lambda(s,t)\,ds\,dt \mid \mathcal{G}_a \right].$$ It remains to show the second statement. Since $N$ is constructed from $\mathcal{N}$, we have $\mathcal{F}_a^N\subseteq \mathcal{G}_a$. Using the tower property of conditional expectation, for $0\leq a<b$ and $B\in\mathcal{B}(S)$, $$ \mathbb{E}\{N(B\times(a,b])\mid \mathcal{F}_a^N\} = \mathbb{E} \left[ \mathbb{E}\{N(B\times(a,b])\mid \mathcal{G}_a\} \mid \mathcal{F}_a^N \right]. $$ By the first part of the lemma, $$ \mathbb{E}\{N(B\times(a,b])\mid \mathcal{G}_a\} = \mathbb{E} \left[\int_a^b \int_B \lambda(s,t)\,ds\,dt \mid \mathcal{G}_a \right]. $$ Hence $$ \mathbb{E}\{N(B\times(a,b])\mid \mathcal{F}_a^N\} = \mathbb{E} \left[ \mathbb{E} \left[ \int_a^b \int_B \lambda(s,t)\,ds\,dt \mid \mathcal{G}_a \right] \mid \mathcal{F}_a^N \right]. $$ Applying the tower property again gives $$ \mathbb{E}\{N(B\times(a,b])\mid \mathcal{F}_a^N\} = \mathbb{E} \left[ \int_a^b \int_B \lambda(s,t)\,ds\,dt \mid \mathcal{F}_a^N \right]. $$ Therefore $\lambda(s,t)$ is the conditional intensity of $N$ with respect to the natural filtration $(\mathcal{F}_t^N)_{t\geq0}$. This step moves from the larger information structure $\mathcal{G}_a$, which contains the background Poisson information and the initial condition, to the smaller natural filtration $\mathcal{F}_a^N$, which contains only the history of accepted events up to time $a$. It shows that, when $\lambda(s,t)$ is predictable from the accepted process itself, the same function is also the intensity with respect to the natural history of $N$.
\end{proof}

\medskip

\paragraph{Lemma 4.3 (Construction).}
Let $S\subset \mathbb{R}^2$ be bounded, and let $y(s,t)$ be a fixed realisation of the latent Gaussian field $Y(s,t)$. Suppose that $$ b(s,t)=\mu(s,t)+y(s,t) $$ is Borel measurable and bounded on $S\times[0,T]$ for every finite $T>0$. Let $N^0$ be an initial spatio-temporal point process supported on $S\times(-\infty,0]$, independent of the background Poisson random measure $\mathcal{N}$. Represent the initial process $N^0$ by the points $\{(S_i^0,T_i^0):i\leq 0\}, \: T_i^0\leq 0.$ For $t>0$, define the initial history by $$ \mathcal{H}_t^{(0)} = \{(S_i^0,T_i^0):i\leq 0,\ T_i^0<t\}.$$ Assume the initial history-dependent part $$D_0(s,t) = D(s,t\mid\mathcal{H}_t^{(0)})$$ is finite and bounded on $S\times[0,T]$ for every finite $T>0$. For a history $\mathcal{H}_t=\{(s_i,t_i):t_i<t\}$ for which the following sum is well defined, define the history-dependent component $$D(s,t\mid\mathcal{H}_t) = \alpha \sum_{i:t_i<t} \exp\{-\beta(t-t_i)\} \varphi_{\Sigma}(s-s_i),$$ and the conditional intensity $$ \lambda^y(s,t\mid\mathcal{H}_t) = \exp\{b(s,t)+D(s,t\mid\mathcal{H}_t)\}.$$ Let $\mathcal{N}$ be a homogeneous Poisson random measure on $[0,\infty)\times S\times\mathbb{R}_{\geq0}$ with intensity measure $dt\:ds\:dz$, independent of $N^0$. We construct the future events after time zero recursively. Set $T_0=0$. For $u>0$, define the initial conditional intensity $$ \lambda_0(s,u) = \lambda^y(s,u\mid\mathcal{H}_u^{(0)}) = \exp\{b(s,u)+D_0(s,u)\}. $$ The first future event time is $$ T_1 = \inf \left\{ t>0: \int_{(0,t]\times S\times\mathbb{R}_{\geq0}} \mathbf{1}\{z\leq\lambda_0(s,u)\} \mathcal{N}(du,ds,dz)>0 \right\}. $$ If $T_1<\infty$, let $S_1$ be the spatial coordinate of the corresponding accepted point. Suppose that $(S_1,T_1),\ldots,(S_n,T_n)$ have been constructed, with $0<T_1<\cdots<T_n<\infty.$ For $u>T_n$, define the history $$ \mathcal{H}_u^{(n)} = \mathcal{H}_u^{(0)} \cup \{(S_i,T_i):i=1,\ldots,n,\ T_i<u\}. $$ Since $u>T_n$, all first $n$ future events are included in the history. Hence $$D(s,u\mid\mathcal{H}_u^{(n)}) = D_0(s,u) + \alpha \sum_{i=1}^{n} \exp\{-\beta(u-T_i)\} \varphi_{\Sigma}(s-S_i).$$ Set $$ \lambda_n(s,u) = \lambda^y(s,u\mid\mathcal{H}_u^{(n)}) = \exp\{b(s,u)+D(s,u\mid\mathcal{H}_u^{(n)})\}. $$ The next future event time is $$ T_{n+1} = \inf \left\{ t>T_n: \int_{(T_n,t]\times S\times\mathbb{R}_{\geq0}} \mathbf{1}\{z\leq\lambda_n(s,u)\} \mathcal{N}(du,ds,dz)>0 \right\}. $$ If $T_{n+1}<\infty$, let $S_{n+1}$ be the spatial coordinate of the corresponding accepted point. If $T_{n+1}=\infty$, set $T_k=\infty$ for all $k>n+1$, and the construction stops. Recall that the explosion time is $T_{\infty} = \lim_{n\to\infty}T_n.$ The constructed counting measure is $$ N(A) = N^0(A\cap(S\times(-\infty,0])) + \sum_{i:T_i<T_{\infty}} \mathbf{1}\{(S_i,T_i)\in A\}, \: A\in\mathcal{B}(S\times\mathbb{R}). $$ Then, conditional on $Y(s,t)=y(s,t)$, the constructed process is a spatio-temporal point process up to the explosion time $T_{\infty}$. For $t<T_{\infty}$, its conditional intensity is $$ \lambda^y(s,t\mid\mathcal{H}_t) = \exp\{\mu(s,t)+y(s,t)+D(s,t\mid\mathcal{H}_t)\}. $$

\medskip

\begin{proof}
Before the explosion time, only finitely many future events have been constructed. On each interval $(T_n,T_{n+1})$, the past events consist of the initial history together with the first $n$ future events. Hence the history used by the constructed process is $\mathcal{H}_u^{(n)}$. Therefore, $$ \lambda^y(s,u\mid\mathcal{H}_u) = \lambda_n(s,u), \: u\in(T_n,T_{n+1}). $$ We first check that the recursive construction is locally well defined before explosion. The Gaussian spatial kernel is $$\varphi_{\Sigma}(h) = \frac{1}{2\pi|\Sigma|^{1/2}} \exp\left\{ -\frac{1}{2}h^\top\Sigma^{-1}h \right\}. $$ Its maximum is attained at $h=0$, and hence $$ C_{\Sigma} = \sup_{h\in\mathbb{R}^2}\varphi_{\Sigma}(h) = \varphi_{\Sigma}(0) = \frac{1}{2\pi|\Sigma|^{1/2}} <\infty. $$ Since $\beta>0$ and $u>T_i$ for each previously observed future event, we have $$\exp\{-\beta(u-T_i)\}\leq 1.$$ Hence, for any finite $t>T_n$ and any $u\in[T_n,t]$, $$\left| \alpha \sum_{i=1}^{n} \exp\{-\beta(u-T_i)\} \varphi_{\Sigma}(s-S_i) \right| \leq |\alpha|\,n\,C_{\Sigma}. $$ Since $b(s,u)$ and $D_0(s,u)$ are bounded on $S\times[0,t]$, there exists a finite constant $Q_{n,t}$ such that $$ b(s,u)+D_0(s,u) \leq Q_{n,t}, \: (s,u)\in S\times[T_n,t]. $$ Therefore, we have $$\lambda_n(s,u) \leq \exp\{Q_{n,t}+|\alpha|\,n\,C_{\Sigma}\}. $$ As $S$ is bounded, $$\int_{T_n}^{t}\int_S \lambda_n(s,u)\:ds\:du \leq |S|(t-T_n) \exp\{Q_{n,t}+|\alpha|\,n\,C_{\Sigma}\} <\infty. $$ Thus the accepted region under $\lambda_n$ has finite Poisson measure on every finite interval before the next event time. Because time is continuous, the probability that two background Poisson candidate points have exactly the same time coordinate is zero. Therefore, two accepted candidate points also do not occur at exactly the same time with probability one. Hence, whenever $T_{n+1}<\infty$, there is almost surely a unique accepted point at time $T_{n+1}$, and its spatial coordinate $S_{n+1}$ is well defined. For every $t<T_{\infty}$, the piecewise recursive construction gives $$ N(A\cap(S\times(0,t])) = \int_0^t \int_S \int_{\mathbb{R}_{\geq0}} \mathbf{1}\{(s,u)\in A\} \mathbf{1}\{z\leq\lambda^y(s,u\mid\mathcal{H}_u)\} \mathcal{N}(du,ds,dz). $$ The integrand depends only on events before time $u$, and is therefore predictable with respect to the natural history of the constructed process. By Lemma 4.2, the accepted point process has conditional intensity $$ \lambda^y(s,t\mid\mathcal{H}_t) = \exp\{\mu(s,t)+y(s,t)+D(s,t\mid\mathcal{H}_t)\} $$ for $t<T_{\infty}$. 
\end{proof}

\section{Non-Explosion and Explosion}
We now give a simple sufficient condition for non-explosion ($T_{\infty} = \infty$) in the non-excitatory case, where $\alpha\leq0$. The result is again stated conditional on a realisation of the latent Gaussian field.

\paragraph{Proposition 5.1 (Non-explosion in the non-excitatory case).}
Let $S\subset\mathbb{R}^2$ be bounded, and let $y(s,t)$ be a fixed realisation of the latent Gaussian field. Suppose that $b(s,t)=\mu(s,t)+y(s,t)$ is Borel measurable and bounded on $S\times[0,T]$ for every finite $T>0$. Let the process be constructed by the Poisson embedding construction of Lemma 4.3, with an initial history for which the history-dependent field is well defined. If $\alpha\leq0,$ then the constructed process is non-explosive. That is, $ T_{\infty}=\infty $ almost surely, given the fixed realisation $Y(s,t)=y(s,t)$.

\medskip

\begin{proof}
Since the Gaussian spatial kernel is nonnegative and the temporal decay term is positive, we have $$ \exp\{-\beta(t-t_i)\}>0, \quad  \varphi_{\Sigma}(s-s_i)\geq0.$$ Therefore, when $\alpha\leq0$, every event in the history gives a non-positive contribution to the history-dependent field. Hence, the history-dependent component $$ D(s,t\mid\mathcal{H}_t) = \alpha \sum_{i:t_i<t} \exp\{-\beta(t-t_i)\} \varphi_{\Sigma}(s-s_i) \leq0.$$ It follows that the conditional intensity satisfies $$\lambda^y(s,t\mid\mathcal{H}_t) = \exp\{b(s,t)+D(s,t\mid\mathcal{H}_t)\} \leq \exp\{b(s,t)\}.$$ Fix any finite time $T>0$. Since $b(s,t)$ is bounded on $S\times[0,T]$ and $S$ is bounded, we have $$ \int_0^T\int_S \exp\{b(s,t)\}\,ds\,dt <\infty. $$ Using the same Poisson random measure $\mathcal{N}$ as in the embedding construction, define $$\widetilde{N}_T = \int_0^T \int_S \int_{\mathbb{R}_{\geq0}} \mathbf{1}\{z\leq \exp\{b(s,t)\}\} \mathcal{N}(dt,ds,dz). $$ Then $\widetilde{N}_T$ is a Poisson random variable with finite mean $\int_0^T\int_S \exp\{b(s,t)\}\,ds\,dt.$ Thus $ \widetilde{N}_T<\infty$, then the constructed process is non-explosive conditional on $Y(s,t)=y(s,t)$. Since $$ \lambda^y(s,t\mid\mathcal{H}_t) \leq \exp\{b(s,t)\}, $$ every accepted future point of the proposed process on $S\times(0,T]$ is also contained in the dominating Poisson process. Therefore, $$ N(S\times(0,T]) \leq \widetilde{N}_T <\infty $$ almost surely given $Y(s,t)=y(s,t)$. Then, we need to connect this finite window bound to the explosion time. If $T_{\infty}<\infty$, then infinitely many future events must occur before some finite integer time $k$. Hence $$ \{T_{\infty}<\infty\} \subseteq \bigcup_{k=1}^{\infty} \{N(S\times(0,k])=\infty\}. $$ For each integer $k$, the argument above gives $$ N(S\times(0,k]) \leq \widetilde{N}_k <\infty $$ almost surely conditional on $Y(s,t)=y(s,t)$. Hence $$ \mathbb{P}\left(N(S\times(0,k])=\infty\mid Y(s,t)=y(s,t)\right)=0 $$ for every $k$. Therefore, $$ \mathbb{P}\left(T_{\infty}<\infty\mid Y(s,t)=y(s,t)\right)=0, $$ and hence $$ T_{\infty}=\infty $$ almost surely given $Y(s,t)=y(s,t)$. 
\end{proof}

\medskip

We now show that, in contrast to the non-excitatory case, the excitatory case may lead to explosion. The argument follows the same intuition as the positive near-zero memory case in \cite{bielecki2025loglinear} that repeated events occurring in a sufficiently small time window can increase the log-intensity fast enough to produce infinitely many events in finite time.

\paragraph{Proposition 5.2 (Possible explosion in the excitatory case).}
Let the process be constructed under the assumptions of the construction lemma, and suppose that $\alpha>0$. Let $B\subset S$ be a measurable spatial set with $|B|>0$. Assume that there exists $\delta>0$ such that $m= \inf_{(s,t)\in B\times[0,\delta]} b(s,t) > -\infty.$ Let $c_B = \inf_{s\in B,\:s'\in S} \varphi_{\Sigma}(s-s').$ Since $\varphi_{\Sigma}$ is a positive Gaussian kernel and $S$ is bounded, we have $c_B>0$. Then, conditional on $Y(s,t)=y(s,t)$, the process has positive probability of explosion before time $\delta$. That is, $$ \mathbb{P}\left(T_{\infty}\leq \delta \mid Y(s,t)=y(s,t)\right)>0. $$

\medskip

\begin{proof}
Let $N_S(t)=N(S\times(0,t])$ be the number of events occurring in the spatial region $S$ up to time $t$. Let $\tau_k=T_k$ be the time of the $k$-th future event occurring in $S$, and set $\tau_0=0$. For $s\in B$, $s_i\in S$ and $0<t-t_i\leq\delta$, we have $$\exp\{-\beta(t-t_i)\} \geq \exp\{-\beta\delta\} $$ and the Gaussian kernel $\varphi_{\Sigma}(s-s_i) \geq c_B.$ Then, each previous event in $S$ occurring within the time interval $(0,\delta]$ contributes at least $$ a = \alpha \exp\{-\beta\delta\}c_B > 0 $$ to the history-dependent component at each $s\in B$. On the event that $N_S(t-)=n$ for some $0<t\leq\delta$ with $t<T_{\infty}$, these $n$ events contribute at least $a$, and the initial history contributes non-negatively. Hence $$ D(s,t\mid\mathcal{H}_t) \geq an, \: s\in B.$$ Since $b(s,t)\geq m$ on $B\times[0,\delta]$, it follows that the total conditional rate of events in $S$ satisfies $$\int_S \lambda^y(s,t\mid\mathcal{H}_t)\,ds \geq \int_B \lambda^y(s,t\mid\mathcal{H}_t)\,ds \geq |B|\exp\{m+an\}.$$ We need to show there is a positive probability that the waiting times between events become so small that infinitely many events occur before time $\delta$. Choose $\varepsilon>0$ such that $ \varepsilon \sum_{k=1}^{\infty} \frac{1}{k^2} < \delta. $ Define $$E_n = \bigcap_{k=1}^{n} \left\{ \tau_k-\tau_{k-1} < \frac{\varepsilon}{k^2} \right\}.$$ On $E_n$, the first $n$ waiting times between successive events in $S$ are all sufficiently small. In particular, on $E_{k-1}$, we have $\tau_{k-1}+\varepsilon/k^2\leq\varepsilon\sum_{j=1}^{k}j^{-2}<\delta$. Before the $k$-th event in $S$, there are already $k-1$ future events in $S$. Hence, for $\tau_{k-1}<t\leq\tau_{k-1}+\varepsilon/k^2$ before the next event, the total conditional rate is bounded from below by $$r_k = |B|\exp\{m+a(k-1)\}.$$ Using the Poisson embedding construction, conditional on the history at $\tau_{k-1}$, the probability that the next event in $S$ occurs within time $\varepsilon/k^2$ is at least $1-\exp\{-r_k\varepsilon/k^2\}$. Therefore, averaging over the histories in $E_{k-1}$ gives $$\mathbb{P}\left(\tau_k-\tau_{k-1}<\frac{\varepsilon}{k^2}\mid E_{k-1}\right)\geq 1-\exp\left\{-r_k\frac{\varepsilon}{k^2}\right\}.$$ Substituting $r_k = |B|\exp\{m+a(k-1)\}$ gives $$\mathbb{P}\left(\tau_k-\tau_{k-1}<\frac{\varepsilon}{k^2}\mid E_{k-1}\right)\geq 1-\exp\left\{-\frac{\varepsilon}{k^2}|B|\exp\{m+a(k-1)\}\right\}.$$ Consequently, $$\mathbb{P}\left(\bigcap_{k=1}^{\infty}\left\{\tau_k-\tau_{k-1}<\frac{\varepsilon}{k^2}\right\}\mid Y(s,t)=y(s,t)\right) \geq \prod_{k=1}^{\infty}\left[1-\exp\left\{-\frac{\varepsilon}{k^2}|B|\exp\{m+a(k-1)\}\right\}\right].$$ The infinite product is strictly positive because $$\sum_{k=1}^{\infty}\exp\left\{-\frac{\varepsilon}{k^2}|B|\exp\{m+a(k-1)\}\right\}<\infty.$$ Hence the event $\bigcap_{k=1}^{\infty}\left\{\tau_k-\tau_{k-1}<\frac{\varepsilon}{k^2}\right\}$ has positive conditional probability, that is $$\mathbb{P}\left(\bigcap_{k=1}^{\infty}\left\{\tau_k-\tau_{k-1}<\frac{\varepsilon}{k^2}\right\}\mid Y(s,t)=y(s,t)\right) > 0.$$ Let $A=\bigcap_{k=1}^{\infty}\left\{\tau_k-\tau_{k-1}<\frac{\varepsilon}{k^2}\right\}.$ The argument above shows that $\mathbb{P}(A\mid Y(s,t)=y(s,t))>0$. On $A$, the time of the $n$-th future event in $S$ satisfies $\tau_n = \sum_{k=1}^{n}(\tau_k-\tau_{k-1}) < \sum_{k=1}^{n}\frac{\varepsilon}{k^2}.$ Letting $n\to\infty$ gives $$ \lim_{n\to\infty}\tau_n \leq \sum_{k=1}^{\infty}\frac{\varepsilon}{k^2} < \delta.$$ So on $A$, infinitely many events occur in $S\times(0,\delta]$. Therefore $A\subseteq\{T_{\infty}\leq\delta\}$, and so $$ \mathbb{P}\left(T_{\infty}\leq\delta\mid Y(s,t)=y(s,t)\right) \geq \mathbb{P}(A\mid Y(s,t)=y(s,t)) > 0. $$
\end{proof}

\section{Log-Gaussian Cox-Hawkes Process with Recovery Time}
\paragraph{Definition 6.1 (Log-Gaussian Cox-Hawkes Process with Recovery Time).}
Let $S\subset\mathbb{R}^2$ be bounded, and let $Y(s,t)=y(s,t)$ be a fixed realisation of the latent Gaussian field. Let $\rho>0$ be a fixed recovery time. A spatio-temporal point process $N$ on $S\times[0,\infty)$ with explosion time $T_{\infty}\in(0,\infty]$ is called a log-Gaussian Cox-Hawkes process with recovery time $\rho$ if its conditional intensity is $$ \lambda_{\rho}^y(s,t\mid\mathcal{H}_t) = \exp\{\mu(s,t)+y(s,t)+D_{\rho}(s,t\mid\mathcal{H}_t)\} \mathbf{1}\{t<T_{\infty}\}, $$ where $$ D_{\rho}(s,t\mid\mathcal{H}_t) = \alpha \sum_{i:t_i<t-\rho} \exp\{-\beta(t-t_i)\} \varphi_{\Sigma}(s-s_i). $$ Here $\rho$ represents a recovery time: an event at time $t_i$ has no effect on the intensity during the interval $(t_i,t_i+\rho]$, and it can affect the future intensity only after time $t_i+\rho$.

\medskip

\paragraph{Proposition 6.2 (Existence up to explosion with recovery time).}
Let $S\subset\mathbb{R}^2$ be bounded, and let $y(s,t)$ be a fixed realisation of the latent Gaussian field. Suppose that $b(s,t)=\mu(s,t)+y(s,t)$ is Borel measurable and bounded on $S\times[0,T]$ for every finite $T>0$. Let $\rho>0$ be fixed, and define
$$D_{\rho}(s,t\mid\mathcal{H}_t)=\alpha\sum_{i:t_i<t-\rho}\exp\{-\beta(t-t_i)\}\varphi_{\Sigma}(s-s_i).$$
Assume that the initial history contribution $D_{\rho,0}(s,t)$ is finite and bounded on $S\times[0,T]$ for every finite $T>0$. Then, conditional on $Y(s,t)=y(s,t)$, the Poisson embedding construction defines a log-Gaussian Cox-Hawkes process with recovery time up to its explosion time $T_{\infty}$. For $t<T_{\infty}$, its conditional intensity is
$$\lambda_{\rho}^y(s,t\mid\mathcal{H}_t)=\exp\{\mu(s,t)+y(s,t)+D_{\rho}(s,t\mid\mathcal{H}_t)\}.$$

\medskip

\begin{proof}
This follows from the same Poisson embedding construction as Lemma 4.3, with $D$ replaced by $D_{\rho}$. Indeed, before explosion, the history contains only finitely many future events. Since $\varphi_{\Sigma}$ is bounded and $b(s,t)$ and $D_{\rho,0}(s,t)$ are bounded on $S\times[0,T]$ for every finite $T>0$, the intensity is bounded over $S$ on every finite time interval before the next accepted event. Hence the recursive construction is locally well defined. The resulting process satisfies the Poisson embedding representation with predictable acceptance function $\lambda_{\rho}^y(s,t\mid\mathcal{H}_t)$. By Lemma 4.2, this accepted process has conditional intensity $\lambda_{\rho}^y(s,t\mid\mathcal{H}_t)$ for $t<T_{\infty}$.
\end{proof}

\medskip

\paragraph{Proposition 6.3 (Non-explosion with recovery time).}
Under the assumptions above, the log-Gaussian Cox-Hawkes process with recovery time $\rho>0$ is non-explosive for any $\alpha\in\mathbb{R}$. That is, $T_{\infty}=\infty$ a.s., conditional on $Y(s,t)=y(s,t)$.

\medskip

\begin{proof}
Events cannot affect the intensity during the first $\rho$ units of time after they occur. Fix an interval $I_j=(j\rho,(j+1)\rho],\: j=0,1,2,\ldots.$ For $t\in I_j$, any event occurring after time $j\rho$ satisfies $t_i>j\rho\geq t-\rho$, and therefore it is not included in the sum defining $D_{\rho}(s,t\mid\mathcal{H}_t)$. Hence events born inside $I_j$ cannot increase the intensity inside the same interval. We prove by induction that only finitely many future events occur in each interval $I_j$. For $j=0$, the intensity on $I_0=(0,\rho]$ depends only on the initial history through $D_{\rho,0}(s,t)$. Since $b(s,t)$ and $D_{\rho,0}(s,t)$ are bounded on $S\times[0,\rho]$, there exists a finite constant $M_0$ such that $$\lambda_{\rho}^y(s,t\mid\mathcal{H}_t)\leq \exp\{M_0\},\: (s,t)\in S\times I_0.$$ Since $S$ is bounded, $$\int_{I_0}\int_S \lambda_{\rho}^y(s,t\mid\mathcal{H}_t)\,ds\,dt\leq |S|\rho\exp\{M_0\}<\infty.$$ Thus only finitely many events occur in $S\times I_0$ a.s.. Now suppose that only finitely many events have occurred in $S\times(0,j\rho]$. For $t\in I_j$, the recovery time condition implies that $D_{\rho}(s,t\mid\mathcal{H}_t)$ can depend only on the initial history and on the finitely many future events before time $j\rho$. Since $\varphi_{\Sigma}$ is bounded and $b(s,t)$ and $D_{\rho,0}(s,t)$ are bounded on $S\times[0,(j+1)\rho]$, there exists a finite constant $M_j$ such that $$\lambda_{\rho}^y(s,t\mid\mathcal{H}_t)\leq \exp\{M_j\},\: (s,t)\in S\times I_j.$$ Therefore, $$\int_{I_j}\int_S \lambda_{\rho}^y(s,t\mid\mathcal{H}_t)\,ds\,dt\leq |S|\rho\exp\{M_j\}<\infty.$$ Hence only finitely many events occur in $S\times I_j$ a.s.. By induction, each interval $I_j$ contains only finitely many future events. Now fix any finite time $T>0$. Choose $J$ such that $T\leq J\rho$. Then $$N(S\times(0,T])\leq \sum_{j=0}^{J-1}N(S\times I_j)<\infty$$ a.s., the process has only finitely many events on every finite time interval. Therefore, $$\mathbb{P}\left(T_{\infty}<\infty\mid Y(s,t)=y(s,t)\right)=0,$$ and so $T_{\infty}=\infty$ a.s., conditional on $Y(s,t)=y(s,t)$. 
\end{proof}

\section{Physical Interpretation of the Log-Gaussian Cox-Hawkes Process with Recovery Time for Stochastic Star Formation}
Motivated by stochastic self-propagating star formation models and recent connections between SSPSF and spatio-temporal point processes, we propose a new spatio-temporal log-Gaussian Cox-Hawkes process for stochastic star formation. Recall that the log-Gaussian Cox-Hawkes process with recovery time provides a continuous stochastic formulation of self-propagating star formation. The model is specified through the conditional intensity $$ \lambda_\rho^y(s,t\mid \mathcal{H}_t) = \exp \{\mu(s,t)+y(s,t)+D_\rho(s,t\mid \mathcal{H}_t)\}, $$ where $$ D_\rho(s,t\mid \mathcal{H}_t) = \alpha \sum_{i:t_i<t-\rho} \exp\{-\beta(t-t_i)\}\varphi_\Sigma(s-s_i).$$ Here, $u=t-t_i$ is the age of a previous star formation event, $h=s-s_i$ is the spatial displacement from that star formation event and $\rho$ is the recovery time. The mean trend $\mu(s,t)$ represents the deterministic background tendency for star formation. Physically, this component may describe large-scale galactic structure across the galactic disk. It acts as the role of spontaneous star formation in the classical SSPSF framework, but allows the spontaneous component to vary continuously over space and time. This is important because the background tendency for star formation in a galaxy is expected to vary continuously across space and over time, rather than remain fixed at a constant probability. The latent field $y(s,t)$ represents unobserved galactic background variation. Star formation is affected by physical conditions that may not be fully observed, such as local gas density, metallicity, pressure, magnetic fields, or dynamical structure. The log-Gaussian Cox component allows these hidden environmental effects to generate additional spatio-temporal clustering beyond that explained by the deterministic background mean field. The history-dependent field $D_\rho(s,t\mid \mathcal{H}_t)$ represents the self-propagation of star-formation activity. It describes the effect of previous star-formation events on the future log-intensity of star formation nearby in space and time. The parameter $\alpha$ controls the sign and strength of this history effect. When $\alpha>0$, previous events increase the future local intensity, corresponding to positive feedback or triggered star formation. When $\alpha<0$, previous events decrease the future local intensity, corresponding to local suppression caused by feedback that temporarily reduces the availability or suitability of nearby gas for further star formation. The parameter $\beta>0$ controls the temporal decay of the history effect in star formation. Larger values of $\beta$ imply that the influence of a previous event decreases more quickly, while smaller values imply a longer-lasting influence. The spatial kernel $\varphi_\Sigma(s-s_i)$ controls the spatial range and shape of the propagation effect. The covariance matrix $\Sigma$ determines whether this influence is isotropic or anisotropic, and therefore allows the propagation of star formation to spread differently in different spatial directions. The recovery time $\rho>0$ should be interpreted as a delay before a previous event can affect the future intensity. An event occurring at time $t_i$ does not contribute to the history-dependent field during the interval $(t_i,t_i+\rho]$. It can only affect the future intensity after its age exceeds $\rho$. It represents a finite feedback delay or source-side recovery time, rather than a cell refractory period. Physically, this delay represents the time required for massive stars to evolve, for supernova feedback or for feedback from a star-forming region to become capable of influencing neighbouring gas. Mathematically, the delayed interaction prevents events from immediately increasing the intensity after they occur, thereby avoiding instantaneous self-triggering cascades and supporting the non-explosion property of the model.

\section{Extensions with Interaction Coefficient Fields}
\label{sec:interaction-coefficient-fields}
The preceding model uses a constant coefficient $\alpha$ to describe the direction and magnitude of the history effect. We now introduce spatio-temporal interaction coefficient fields while retaining the background $\mu(s,t)+Y(s,t)$, the temporal decay parameter $\beta$ and the Gaussian spatial kernel $\varphi_\Sigma$ defined above. The conditional intensity retains the form $$\lambda(s,t\mid\mathcal{H}_t)=\exp\{\mu(s,t)+Y(s,t)+D(s,t\mid\mathcal{H}_t)\},$$ with the dynamic field $D$ specified below. The conditional intensities are considered before the explosion time $T_\infty$. Unless stated otherwise, $\alpha_R$ and $\alpha_S$ are unknown deterministic real-valued functions on $\mathcal{S}\times[0,\infty)$, assumed Borel measurable and bounded on $\mathcal{S}\times[0,T]$ for every finite $T>0$. Replacing $D$ by $D_R$, $D_S$, $D_{RS}$ or $D_{RS}^{\mathrm{add}}$ defines the respective conditional intensities $\lambda_R$, $\lambda_S$, $\lambda_{RS}$ and $\lambda_{RS}^{\mathrm{add}}$. 

\subsection{Response Field Model}
The constant coefficient $\alpha$ is replaced by a spatio-temporal response field $\alpha_R(s,t)$, allowing the direction and strength of the history effect to vary across space and time. The dynamic field is defined as $$D_R(s,t\mid\mathcal{H}_t)=\alpha_R(s,t)\sum_{i:t_i<t}\exp\{-\beta(t-t_i)\}\varphi_\Sigma(s-s_i).$$

The term response refers to the sensitivity of the local log-intensity to the weighted sum of past events appearing in this expression. The field $\alpha_R(s,t)$ is evaluated at the location and time where the intensity is modelled and controls the direction and strength of this dependence. Positive values correspond to excitation and negative values to inhibition, while larger absolute values produce a stronger history effect for the same weighted sum of past events.

\subsection{Source Field Model}
Each past event $(s_i,t_i)$ acts as a source of influence on the future log-intensity and is assigned a signed coefficient $\alpha_S(s_i,t_i)$ by evaluating the spatio-temporal source field $\alpha_S(s,t)$ at its location and occurrence time. This coefficient remains fixed for that event, while the temporal and spatial kernels determine how its influence decays. Different source events may thus contribute with opposite signs at the same $(s,t)$. Summing their contributions gives $$D_S(s,t\mid\mathcal{H}_t)=\sum_{i:t_i<t}\alpha_S(s_i,t_i)\exp\{-\beta(t-t_i)\}\varphi_\Sigma(s-s_i).$$

\subsection{Multiplicative Source-Response Model}
The multiplicative source-response model combines the two fields through their product, giving $$D_{RS}(s,t\mid\mathcal{H}_t)=\alpha_R(s,t)\sum_{i:t_i<t}\alpha_S(s_i,t_i)\exp\{-\beta(t-t_i)\}\varphi_\Sigma(s-s_i).$$ The sign of each historical contribution is determined by $\alpha_R(s,t)\alpha_S(s_i,t_i)$, equal non-zero signs produce excitation, whereas opposite signs produce inhibition. The interaction amplitude is separable in the source coordinates $(s_i,t_i)$ and the response coordinates $(s,t)$. The two fields enter the dynamic field only through their product. Multiplying $\alpha_R$ by a non-zero constant $c$ and dividing $\alpha_S$ by the same constant leaves this product unchanged, since $$[c\alpha_R(s,t)][c^{-1}\alpha_S(s_i,t_i)]=\alpha_R(s,t)\alpha_S(s_i,t_i).$$ These different pairs of fields therefore produce the same conditional intensity. In particular, taking $c=-1$ reverses the signs of both fields without changing the model. The individual scales and signs of the two fields cannot be distinguished from the conditional intensity alone. A scale normalization and a sign convention are needed to remove this ambiguity when interpreting the fields separately.

\subsection{Additive Source-Response Model}
The additive source-response model specifies $$D_{RS}^{\mathrm{add}}(s,t\mid\mathcal{H}_t)=\sum_{i:t_i<t}\{\alpha_R(s,t)+\alpha_S(s_i,t_i)\}\exp\{-\beta(t-t_i)\}\varphi_\Sigma(s-s_i).$$ The source and response fields contribute additively to the interaction amplitude. The sign of each historical contribution is determined by $\alpha_R(s,t)+\alpha_S(s_i,t_i)$, allowing the two contributions to reinforce or offset one another. The addition takes place within the history-dependent component of the log-intensity. Adding a constant $c\in\mathbb{R}$ to $\alpha_R$ and subtracting the same constant from $\alpha_S$ leaves their sum unchanged. These different pairs of fields therefore produce the same conditional intensity, so the constant components of the two fields cannot be distinguished without an additional constraint. 

\subsection{Random Interaction Field Models}
Each interaction coefficient field may be specified as a latent random field. In the response and source field models, this applies to $\alpha_R$ and $\alpha_S$, respectively. In the multiplicative and additive source-response models, one field may be random while the other remains deterministic, or both fields may be random. When both are random, their joint distribution may specify independence or allow dependence between them. Conditional on $Y$ and the random interaction fields, the dynamic-field definitions above apply without modification and are evaluated at the realised field values.

\subsection{Special Cases and Model Reductions}
The response and source field models recover the original constant-coefficient model when $\alpha_R\equiv\alpha$ and $\alpha_S\equiv\alpha$. With constant fields $\alpha_R\equiv a_R$ and $\alpha_S\equiv a_S$, the multiplicative and additive source-response models recover it with $\alpha=a_Ra_S$ and $\alpha=a_R+a_S$. The multiplicative source-response model reduces to the response field model when $\alpha_S\equiv1$ and to the source field model when $\alpha_R\equiv1$. The corresponding restrictions for the additive source-response model are $\alpha_S\equiv0$ and $\alpha_R\equiv0$. Random interaction-field models reduce to the corresponding deterministic models when the interaction fields equal deterministic functions almost surely.

All four models reduce to a log-Gaussian Cox process when $D\equiv0$. This occurs with a zero interaction field in either single-field model, either field identically zero in the multiplicative model, or opposite constant fields $\alpha_R\equiv c$ and $\alpha_S\equiv-c$ in the additive model. Setting $Y\equiv0$ gives a log-linear Hawkes-type form, conditional on any remaining random interaction fields.

\section{Discussion}
This work developed a unified spatio-temporal point-process framework for stochastic star formation. First, we constructed a conditional Poisson representation of the SSPSF update law. This connection provides a statistical interpretation of the SSPSF mechanism and shows how spontaneous star formation and triggered star formation can be understood in terms of background intensity and history-dependent triggering. Building on this connection and motivated by the Cox-Hawkes framework of \cite{miscouridou2022cox}, we then introduced a new spatio-temporal log-Gaussian Cox-Hawkes process with signed history dependence. The model places the deterministic mean field, latent Gaussian field, and history-dependent interaction field inside a single log-intensity. This log-scale construction allows past events to have either excitatory or inhibitory effects while keeping the conditional intensity positive. In this way, the proposed model extends beyond additive Cox-Hawkes formulations and provides a new interpretable point-process model for stochastic star formation. In particular, it can represent latent clustering, self-excitation, local inhibition, and event-driven propagation within a single continuous spatio-temporal framework. 

We also introduced a recovery time version of the log-Gaussian Cox-Hawkes process as a stochastic star-formation model with a direct physical interpretation. The recovery time prevents recently formed events from immediately affecting the intensity and reflects the delay required before an event can contribute to subsequent star formation. This formulation is therefore not simply a continuous version of the classical SSPSF model, but a new stochastic point-process framework for understanding, simulating, modelling, and performing inference for star-formation activity in galaxies. We also introduced extensions in which the constant interaction coefficient is replaced by spatio-temporal response and source fields, combined multiplicatively or additively when both are present. These fields may be deterministic or random, allowing variation in the direction and strength of history dependence across space and time.

Several methodological and theoretical questions remain open. On the theoretical side, following \cite{bielecki2025loglinear}, the results for the constant-coefficient formulation establish existence up to explosion, non-explosion in the non-excitatory case, possible explosion in the excitatory case, and non-explosion for the recovery-time model. A full stability theory remains to be developed, and the theoretical properties of the interaction-field extensions require separate analysis. The interaction between the latent Gaussian field and the history-dependent component may also raise further questions about parameter identifiability and asymptotic behaviour. On the methodological side, the conditional log-likelihood contains a spatio-temporal integral involving both the latent Gaussian field and the history-dependent component, which is generally not available in closed form. As a result, inference is likely to require numerical approximation together with treatment of the latent field. Possible approaches include discretisation-based likelihood approximation, Monte Carlo likelihood methods \citep[see e.g.,][]{geyer1992constrained, geyer1994convergence, geyer1994simulation, christensen2004monte}, and Bayesian inference \citep{rasmussen2013bayesian, teng2017bayesian, adams2009tractable, rue2009approximate, lindgren2011explicit}. Future work will focus on stability theory, parameter identifiability, estimation methods, simulation studies, and applications to simulated or observed galaxy data.

\section*{Acknowledgements}
The author gratefully acknowledges the helpful exchanges on this work with colleagues from their time as a graduate student at the University of Melbourne.

\section*{Declarations}
The author has no competing interests to declare. This manuscript was prepared without financial support. No datasets were generated or analysed in this work.

\appendix
\section{Classical SSPSF Model and Its Point-Process Interpretation}
\label{app:sspsf-stpp}
In this section, we build a bridge between the STPP and SSPSF frameworks. The notation and formulation are self-contained. We construct a conditional Poisson representation of the SSPSF update rule under the stated assumptions. 

\subsection{The Classical Stochastic Self-Propagating Star Formation Model}
\label{sec:G&S1978}
In this work, we focus mainly on the classical Stochastic Self-Propagating Star Formation (SSPSF) model. This model represents a galactic disk as a two-dimensional array of concentric rings, each subdivided azimuthally into equal-area cells, which represent regions of the galaxy where stars can form. The disk rotates differentially according to a prescribed rotation curve, such that the set of nearest neighbours of a given cell changes over time as the rings shear past one another. Each cell of the disk can contain at most one star at a given time step, where a star refers to a star-forming region such as a young stellar cluster that is capable of producing massive stars that can trigger further star formation in surrounding regions. A supernova denotes a cell whose massive stars produce feedback strong enough to initiate star formation in its surrounding cells.  This discrete, probabilistic stochastic star formation framework couples local star formation propagation to the global galaxy disk, generating persistent spiral structure without invoking any density wave. The SSPSF formulation considered here evolves according to the following rules:
\begin{enumerate}
    \item Initialisation of the model: randomly assign $K$ cells with stars before simulation. Every star becomes a supernova in the very next time step after it is formed.
    \item Triggered star: Each surrounding supernova cell independently triggers star formation in an inactive cell with probability $P_t(t_a)$, independently of spontaneous star formation, where $0 \leq P_t \leq 1$ and $t_r$ is the recovery period. 
    \begin{align*}
        P_t(t_a) =
        \begin{cases}
            P_{t} \dfrac{ t_{a}}{t_r}, & 0 \le  t_{a} < t_r, \\
            P_{t}, & t_{a} \ge t_r,
        \end{cases}
        \label{eq:refractory}
    \end{align*}
    where $t_{a}$ is the age in time steps since the last star formed in the cell.
    \item Spontaneous star: Any inactive cell can spontaneously form a star with a spontaneous probability ($0 \leq P_{s} \leq 1$), independent of its neighbours. 
    \item Aging and Triggering: the region ages by one time step at each iteration after a star was born. Massive stars trigger surrounding cells in the time step immediately after their creation.
    \item Rotation: After each time step, rings are rotated according to the pre-designed galaxy's rotation curve. 
\end{enumerate}
Define $C_{i,j}(t)$ as the activity state of the cell in the ring $i$ and the azimuth $j$ at time step $t$ ($0$ = inactive, $1$ = newly active). Let $\mathcal{F}_t$ denote the discrete history through step $t$. If $C_{i,j}(t) = 0$, the probability of a new star forming in this region is $$\Pr(C_{i,j}(t+1) = 1\mid C_{i,j}(t)=0,\mathcal{F}_t) = 1 - (1 - P_{s}) \prod_{n \in N_{i,j}(t)} ( 1 - P_{t}(A_{i,j}(t))),$$ where $N_{i,j}(t)$ is the set of neighbouring cells with $C_n(t)=1$ that can trigger the next update, and $A_{i,j}(t)$ is the number of time steps since the last star-formation event in cell $(i,j)$. Set $A_{i,j}(t+1)=0$ after a new event and otherwise set $A_{i,j}(t+1)=A_{i,j}(t)+1$, including when the cell is inactive. Cells with no previous event are initialised with $A_{i,j}(0)\geq t_r$. A cell with $C_{i,j}(t)=1$ becomes inactive after its one-step triggering phase, so $C_{i,j}(t+1)=0$. The probability of triggered star formation denoted $P_{t}$ and the probability of spontaneous star formation $P_{s}$ are two main parameters used to control the star formation process.

\subsection{A conditional Poisson representation of the SSPSF update law}
Discretize space into equal-area cells $\{S_{i,j}\}$ (ring $i$, azimuth $j$) and time into steps of length $\Delta\tau>0$, and denote discrete times as $t_k=k\,\Delta\tau$. Each cell has positive finite area and finitely many neighbours. Let $\mathcal{F}_k$ denote the discrete history through step $k$, including the cell states and ages at $t_k$. Here, $C_{i,j}(t_k)=0$ denotes an inactive cell and $C_{i,j}(t_k)=1$ denotes a newly active cell. The age $A_{i,j}(t_k)$ and the recovery period $t_r>0$ are measured in time steps. Let $\mathcal{N}_{i,j}(t_k)$ be the set of neighbouring cells with $C_{m,n}(t_k)=1$ that can trigger the next update. The neighbour sets and receiving-cell ages used in the update are held fixed during each step. We assume that, conditional on $\mathcal{F}_k$, spontaneous formation and triggering by individual active neighbours are mutually independent, both within and across receiving cells. Newly formed stars contribute to triggering only in the next update. For $0\leq P_s<1$ and $0\leq P_t<1$, define the background and individual trigger contributions by \begin{align*} \Lambda^{(1)}_{i,j}(t_k) &= -\log(1-P_s),\\ \Lambda^{(2)}_{i,j\leftarrow m,n}(t_k) &= -\log\{1-P_t(A_{i,j}(t_k))\}, \quad S_{m,n}\in\mathcal{N}_{i,j}(t_k), \end{align*} where $P_t(A_{i,j}(t_k))$ is the age-dependent triggering probability defined above. All other trigger contributions are zero. Set \begin{equation} \Lambda_{i,j}(t_k) =\Lambda^{(1)}_{i,j}(t_k) +\sum_{S_{m,n}\in\mathcal{N}_{i,j}(t_k)} \Lambda^{(2)}_{i,j\leftarrow m,n}(t_k). \label{eq:Lambdasum} \end{equation} Conditional on $\mathcal{F}_k$, generate independent Poisson proposal counts $$ Q_{i,j}(t_k)\mid\mathcal{F}_k \sim\mathrm{Poisson}\{\Lambda_{i,j}(t_k)\}. $$ Equivalently, conditional on $\mathcal{F}_k$, these proposals can be generated by a Poisson point process with cell-wise constant intensity $$ \lambda_k(t,\mathbf{s}) =\frac{\Lambda_{i,j}(t_k)}{|S_{i,j}|\Delta\tau}, \quad \mathbf{s}\in S_{i,j},\quad t_k<t\leq t_{k+1}. $$ Thus, $$ \int_{t_k}^{t_{k+1}}\int_{S_{i,j}} \lambda_k(t,\mathbf{s})\,d\mathbf{s}\,dt =\Lambda_{i,j}(t_k). $$ The Poisson law is part of this conditional construction. In particular, \begin{align} \Pr(Q_{i,j}(t_k)\geq1\mid\mathcal{F}_k) =1-\exp\{-\Lambda_{i,j}(t_k)\}. \label{eq:atleastone} \end{align} For an inactive cell, retain a single star-formation event if at least one proposal occurs. An active cell becomes inactive after its one-step triggering phase, and proposals in that cell are ignored. Hence, $$ C_{i,j}(t_{k+1}) =\mathbf{1}\{C_{i,j}(t_k)=0\} \mathbf{1}\{Q_{i,j}(t_k)\geq1\}. $$ Only the retained event contributes to the subsequent state and age updates; multiple proposals do not produce additional stars or additional triggering contributions. Using \eqref{eq:atleastone} and \eqref{eq:Lambdasum}, for an inactive cell we obtain \begin{align*} \Pr(C_{i,j}(t_{k+1})=1 \mid C_{i,j}(t_k)=0,\mathcal{F}_k) &=1-\exp\{-\Lambda^{(1)}_{i,j}(t_k)\} \prod_{S_{m,n}\in\mathcal{N}_{i,j}(t_k)} \exp\{-\Lambda^{(2)}_{i,j\leftarrow m,n}(t_k)\}\\ &\quad=1-(1-P_s) \prod_{S_{m,n}\in\mathcal{N}_{i,j}(t_k)} \{1-P_t(A_{i,j}(t_k))\}. \end{align*} This is exactly the SSPSF update probability specified above. The recovery rule is incorporated through $$ P_t(t_a)=r(t_a,t_r)P_t, \quad r(t_a,t_r)=\min\left(1,\frac{t_a}{t_r}\right). $$ Consequently, the required contribution from each active neighbour is $$ \Lambda^{(2)}_{i,j\leftarrow m,n}(t_k) =-\log\{1-r(A_{i,j}(t_k),t_r)P_t\}. $$ After a retained event, set $A_{i,j}(t_{k+1})=0$; otherwise, set $A_{i,j}(t_{k+1})=A_{i,j}(t_k)+1$. Cells with no previous event are initialised with $A_{i,j}(0)\geq t_r$. Boundary probabilities equal to one are interpreted as limits of the binary update probabilities; an effective probability of one requires an infinite Poisson mean. Differential rotation is incorporated by updating the neighbour sets between steps according to the prescribed rotation curve, with states and ages remaining associated with their respective cells. Under the stated independence and update assumptions, this construction reproduces the one-step SSPSF transition law, including recovery and differential rotation.

\bibliographystyle{biorefs}
\bibliography{refs}

\end{document}